\documentclass[aps,prl,twocolumn,showpacs,10pt,superscriptaddress,preprintnumbers]{revtex4-1}
\usepackage{amsmath}
\usepackage{physics}
\usepackage{graphicx}
\usepackage{amssymb}
\usepackage[colorlinks,citecolor=red]{hyperref}

\usepackage{color}    
\usepackage{multirow}
\usepackage{soul}

\newcommand{\beq}{\begin{equation}}
\newcommand{\eeq}{\end{equation}}
\newcommand{\bea}{\begin{eqnarray}}
\newcommand{\eea}{\end{eqnarray}}
\newcommand{\nn}{\nonumber}

\usepackage{ulem}

\begin{document}

\preprint{
	{\vbox {
			\hbox{\bf LA-UR-21-31134}
			\hbox{\bf MSUHEP-21-034}
}}}
\vspace*{0.2cm}

\title{Jet Charge: A new tool to probe the anomalous $Zb\bar{b}$ couplings at the EIC}

\author{Hai Tao Li}
\email{haitao.li@sdu.edu.cn}
\affiliation{School of Physics, Shandong University, Jinan, Shandong 250100, China}

\author{Bin Yan}
\email{yanbin@ihep.ac.cn}
\affiliation{Institute of High Energy Physics, Chinese Academy of Sciences, Beijing 100049, China}
\affiliation{Theoretical Division, Group T-2, MS B283, Los Alamos National Laboratory, P.O. Box 1663, Los Alamos, NM 87545, USA}

\author{C.-P. Yuan}
\email{yuan@pa.msu.edu}
\affiliation{Department of Physics and Astronomy,
Michigan State University, East Lansing, MI 48824, USA}

\date{\today}

\begin{abstract}
We propose to probe the $Zb\bar{b}$ interactions 
at the Electron-Ion Collider (EIC) by 
utilizing the average jet charge weighted single-spin asymmetry $A_e^{bQ}$, which is induced by different cross sections of a left-handed and right-handed electron beam scattering off an unpolarized proton beam in the neutral current deeply-inelastic scattering processes with one observed $b$-tagged jet.
This novel observable at the EIC is sensitive to the axial-vector component of the $Zb\bar{b}$ coupling, providing similar information as the $gg\to Zh$ cross section measurement at the high-luminosity Large Hadron Collider, and is complementary to the single-spin asymmetry measurement at the EIC which is sensitive to the vector component of the $Zb\bar{b}$ coupling.
We show that the apparent degeneracy of the allowed $Zb\bar{b}$ coupling in its vector and axial-vector components, implied by the Large Electron-Positron Collider (LEP) and the Stanford Linear Collider (SLC) precision electroweak data, can be broken  by the $A_e^{bQ}$ measurement at the EIC. 
With a large enough integrated luminosity collected at the EIC,  the measurement of  $A_e^{bQ}$ could also resolve the long-standing discrepancy between bottom quark forward-backward asymmetry $A_{\rm FB}^b$ at the LEP and the Standard Model prediction, with  a strong dependence on the $b$-tagging efficiency. 

\end{abstract}

\maketitle

\noindent {\bf Introduction:~}The electroweak precision measurements at the Large Electron-Positron collider (LEP) and the Stanford Linear Collider (SLC) play a crucial role to test the Standard Model (SM) and beyond. 
An interesting observation at the LEP is that the bottom quark forward-backward asymmetry ($A_{\rm FB}^b$), {\it i.e.}, an asymmetry in the production cross sections of bottom quark traveling along or against the electron beam direction, at the $Z$-pole exhibits a long-standing discrepancy with the SM prediction, with a significance around $2.1\sigma$~\cite{ParticleDataGroup:2020ssz}. Such  a startling deviation has been received much attention in the high energy physics community, and consequently, many new physics (NP) models have been proposed in the literature to explain the LEP data~\cite{Choudhury:2001hs,Agashe:2006at,Gori:2015nqa,Liu:2017xmc}. A well known example is that the model with an underlying approximate custodial symmetry, which introduces a sizable right-handed $Zb\bar{b}$ coupling, while keeping the  left-hand $Zb\bar{b}$ coupling about the same as the SM prediction~\cite{Agashe:2006at}. 
Such theory can not only explain the anomaly of $A_{\rm FB}^b$, but also satisfy the limits from the $R_b$ and $A_b$ measurements at the $Z$-pole.  Here, the $R_b$ is defined as the branching fraction of $Z\to b\bar{b}$ to the inclusive hadronic decay of the $Z$ boson at LEP and $A_b$ is the left-right forward-backward asymmetry of $b$ production at the SLC.

However, the $Zb\bar{b}$ couplings can not be fully determined by the global analysis of the LEP and SLC data. It has been shown that there are two possible solutions 
in the parameter space of $Zb\bar{b}$ vector and axial-vector couplings, 
after combining the $A_b$, $R_b$ and $A_{\rm FB}^b$ measurements at the $Z$-pole and off $Z$-pole~\cite{Choudhury:2001hs}.   Breaking the  above-mentioned degeneracy and further pinning down the coupling strength  as well as confirming or excluding the anomaly of $A_{\rm FB}^b$ measurement at the LEP is one of the major tasks of the particle physics at the current and future colliders.
Recently, we demonstrated that the axial-vector component of $Zb\bar{b}$ coupling can be determined by the precision measurement of $gg\to Zh$ production at the Large Hadron Collider (LHC) and the high-luminosity LHC  (HL-LHC)~\cite{Yan:2021veo}, while its vector component can be better constrained by the measurement of  single-spin asymmetry (SSA) in the polarized lepton-hadron collisions, such as at HERA and the upcoming Electron-Ion Collider (EIC)~\cite{Yan:2021htf}.

It was concluded in~\cite{Yan:2021htf} that the SSA at the EIC is only sensitive to the vector component of the $Zb\bar{b}$ coupling. 
To probe its axial-vector component at the EIC calls for a novel idea, that is to measure the average jet charge weighted single-spin asymmetry (WSSA) of the polarized electron-proton cross section in neutral current deeply-inelastic scattering (DIS) processes with one $b$-tagged jet in the final state.
The WSSA, $A_e^{bQ}$, is defined as 
\beq
A_e^{bQ}= 
\frac{\sigma_{b,+}^{\rm Q} - \sigma_{b,-}^{\rm Q}}{\sigma_{b,+}^{\rm Q} + \sigma_{b,-}^{\rm Q}},
\label{eq:asy}
\eeq
where $ \sigma_{b,\pm}^{\rm Q} $ is the average jet charge weighted total inclusive $b$-tagged DIS cross section of 
a right-handed ($+$) or left-handed ($-$) electron beam scattering off an 
unpolarized proton ($p$) beam, i.e,
\beq
 \sigma_{b,\pm}^{\rm Q}=\int dp_T^j\frac{d\sigma_{b,\pm}^{\rm tot}}{dp_T^j}\langle Q_J\rangle_b(p_T^j).
 \label{eq:sigmabq}
\eeq
Here, $\langle Q_J\rangle_b(p_T^j)$ is the average jet charge of a $b$-tagged jet with transverse momentum $p_T^j$, and can be written as 
\beq
\langle Q_J\rangle_b(p_T^j)=\sum_{q=u,d,c,s,b}\left[f_J^q(p_T^j,\epsilon_q^b)-f_J^{\bar{q}}(p_T^j,\epsilon_q^b)\right]\langle Q_J^q\rangle(p_T^j).
\label{eq:avq}
\eeq
Here, $\langle Q_J^q\rangle$ is the average jet charge of the $q$-type jet, and $f_J^q$ represents the fraction of the $q$-type jets which have been (mis-)tagged as the $b$-tagged jets in the final state. $f_J^q$ depends on the tagged jet $p_T^j$ and $\epsilon_q^b$ which is the efficiency of (mis-)tagging a $q$-jet as a $b$-jet.
The minus sign in Eq.~\eqref{eq:avq} is because of the opposite electric charge of quark and its anti-quark.
It has been demonstrated that the jet charge can be used to separate the quark jets from anti-quark jets and to tag the flavor of the quark jets~\cite{Krohn:2012fg,Waalewijn:2012sv,Li:2019dre,Li:2020rqj,Kang:2020fka,Kang:2021ryr}, and has been measured by both the ATLAS and CMS Collaborations~\cite{ATLAS:2015rlw,CMS:2017yer}. In this work, we propose to measure the $b$-tagged jet charge to probe the $Zb\bar{b}$ coupling at the polarized electron-hadron colliders.

A detailed analysis on how the anomalous $Zb\bar{b}$ coupling could affect the value of SSA has been presented in Ref.~\cite{Yan:2021htf}. It showed that the SSA (by setting $\langle Q_J\rangle_b=1$ in Eq.~\eqref{eq:sigmabq}) depends linearly on the vector component of the $Zb\bar{b}$ coupling through the $\gamma-Z$ interference diagram, while the term linearly proportional to the axial-vector component 
is associated with the $F_3$ structure function, which 
involves the convolution of the difference between quark and anti-quark parton distribution functions (PDFs), ($f_q-f_{\bar{q}}$). 
Since bottom ($q=b$) PDF is generated perturbatively through 
DGLAP evolution~\cite{Moch:2004pa,Catani:2004nc}, 
($f_b - f_{\bar{b}}$) must be zero at the leading-order (LO) and next-to-leading order (NLO), though it can be non-vanishing at the NNLO~\cite{Yan:2021htf}.
That explains why the SSA is only sensitive to the vector component of the $Zb\bar{b}$ coupling. In contrast, the WSSA is expected to be sensitive to its axial-vector component. This is due to the insertion of the average jet charge $\langle Q_J\rangle_b$, which depends on the difference between quark and anti-quark cross sections, in the weighted cross section $\sigma_{b,\pm}^Q$, cf.  Eq.~\eqref{eq:sigmabq}.
Below, we detail our analysis to demonstrate that 
the measurement of WSSA can be used to constrain the axial-vector $Zb\bar{b}$ coupling at the EIC, and provide complementary information to the measurements of the $gg \to Zh$ production cross section at the HL-LHC and the SSA at the EIC.

\vspace{3mm}
\noindent {\bf The jet charge:~}
The jet charge is defined as the transverse momentum-weighted sum of the charges of the jet constituents~\cite{Field:1977fa},
\beq
Q_J=\frac{1}{\left(p_{T}^j\right)^\kappa}\sum_{i\in{\rm jet}}Q_i(p_T^i)^\kappa,~\kappa>0,
\eeq
where $p_T^i$ and $Q_i$ are the transverse momentum and electric charge of particle $i$, respectively.  Based on the soft-collinear effective theory~\cite{Bauer:2000yr,Bauer:2001yt,Bauer:2002nz,Beneke:2002ph}, the average of the jet charge of a quark ($q$) jet is found to be ~\cite{Krohn:2012fg,Waalewijn:2012sv}
\begin{align}\label{eq:Qqj}
\langle Q_J^q\rangle&=\frac{\widetilde{J}_{qq}(p_T^j,R,\kappa,\mu)}{J_q(p_T^j,R,\mu)}\widetilde{D}_q^Q(\kappa)\nn\\
&\times {\rm exp}\left[\int_{\mu_0}^\mu\frac{d\mu^\prime}{\mu^\prime}\frac{\alpha_s(\mu^\prime)}{\pi}\tilde{f}_{q\to qg}(\kappa)\right],
\end{align}
where $R$ and $\mu$ are jet  size and factorization scale, respectively. $J_q(p_T^j,R,\mu)$ is a jet function and $\widetilde{J}_{qq}(p_T^j,R,\kappa,\mu)$ is the $(\kappa+1)$-th Mellin moment of the Wilson coefficient for matching the quark fragmenting jet function onto a quark fragmentation function. The NLO calculation of $J_q$ and $\widetilde{J}_{qq}$ can be found in Refs.~\cite{Waalewijn:2012sv,Ellis:2010rwa,Jain:2011xz}. $\tilde{f}_{q\to qg}(\kappa)$ is the $(\kappa+1)$-th Mellin moment of the splitting function $f_{q \to q g}(z)$. The initial scale of the evolution of fragmentation function is set to $\mu_0=1~{\rm GeV}$ for light jets and $\mu_0=m_Q$ for heavy flavor jets. 
Here, $m_Q$ is taken to be the pole mass of the heavy flavor quark.
The only non-perturbative parameter $\widetilde{D}_q^Q(\kappa)$ depends on the parameter $\kappa$ and the flavor of jet,  which can be written as $\widetilde{D}_q^Q(\kappa)=\sum_h Q_h \int_0^1dx x^\kappa D_q^h(x,\mu_0)$. In this work, they are obtained from PYTHIA simulations~\cite{Sjostrand:2007gs}. Note that the non-perturbative parameter in Eq.~(\ref{eq:Qqj}) is independent of the jet $p_T^j$ and jet size $R$. 

\begin{figure}
	\includegraphics[scale=0.23]{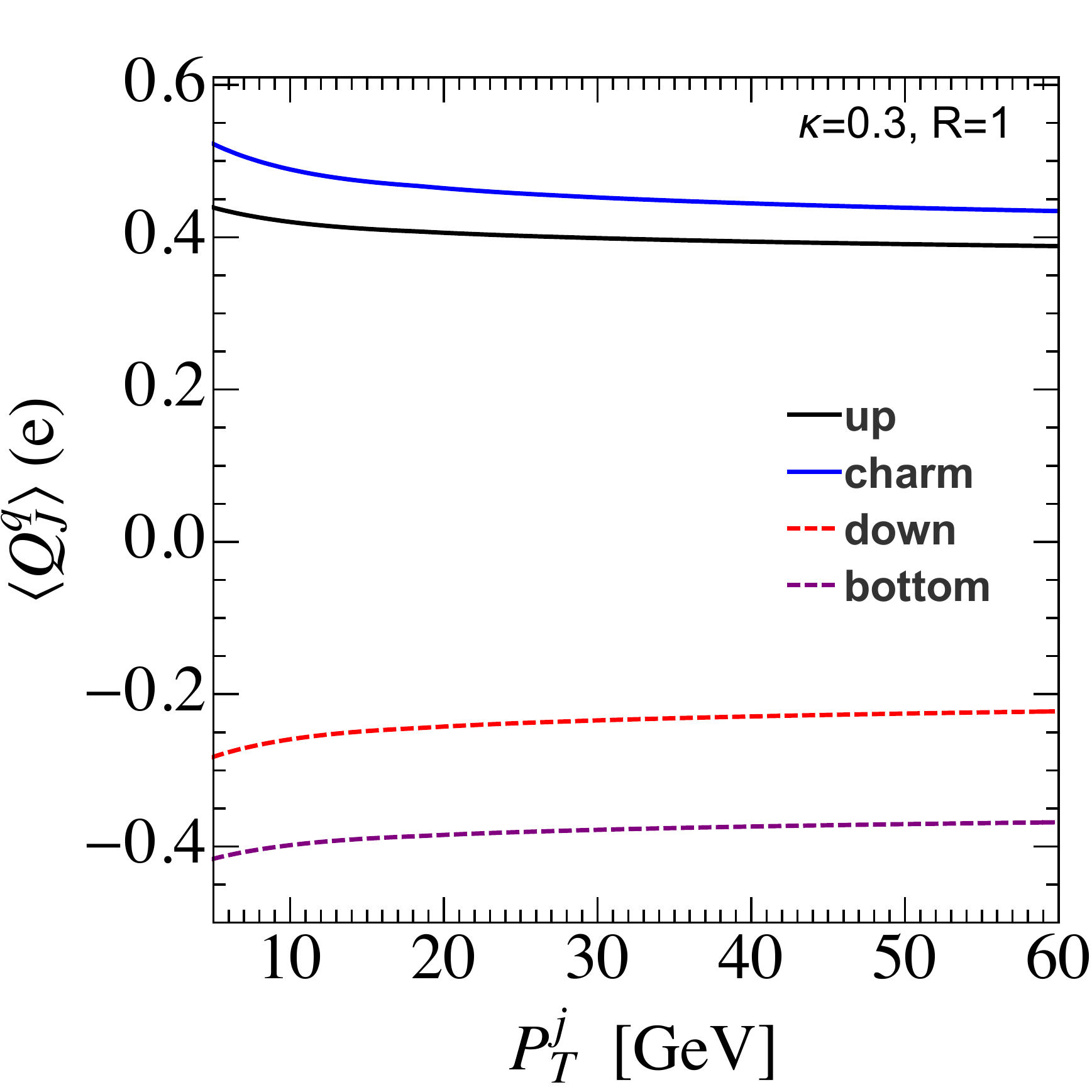}
	\caption{The average jet charge distributions  $\langle Q_J^q\rangle$ for various quark-type jets as a function of jet $p_T^j$, with $\kappa=0.3$. Here, the jet size $R=1$ and the factorization scale $\mu=p_T^jR$.}
\label{Fig:JetQ}
\end{figure}

The average jet charge distribution of various quark jets is depicted in Fig.~\ref{Fig:JetQ}, as a function of jet $p_T^j$ with $\kappa=0.3$, $R=1$ and the factorization scale set to be $\mu=p_T^jR$.  
It shows that the sign of the jet charge is consistent with the sign of the electric charge of the parent parton. 
Furthermore, $\langle Q_J^q\rangle$ is not very sensitive to $p_T^j$, and approaches to a constant value as $p_T^j$ increases. 
For the heavy flavor jet, because the initial scale $\mu_0$ is larger and the fragmentation function $D_{Q}^{h}(z,\mu_0)$ has a much larger value in the large $z$ region~\cite{Braaten:1994bz}, the average jet charge for charm ($c$) or bottom ($b$) jet is larger than the ones for light flavor jets.

\vspace{3mm}
\noindent{\bf DIS cross section:} 
Next, we consider the impact of the non-standard $Zb\bar{b}$ couplings to the WSSA $A_e^{bQ}$ in Eq.~\eqref{eq:asy} at the EIC.
The $Zb\bar{b}$ effective Lagrangian can be parametrized as
\beq
\mathcal{L}_{\rm eff}=\frac{g_W}{2c_W}\bar{b}\gamma_\mu(\kappa_Vg_V^b-\kappa_Ag_A^b\gamma_5)b \, Z_\mu,
\eeq
where $g_W$ is the electroweak gauge coupling and $c_W$ is the cosine of the weak mixing angle $\theta_W$. The couplings $g_V^b=-1/2+2/3s_W^2$ and $g_A^b=-1/2$ with $s_W\equiv\sin\theta_W$ are the vector and axial-vector components of the $Zb\bar{b}$ coupling in the SM, respectively.  We introduce the parameters $\kappa_{V,A}$ to include the possible NP effects for the $Zb\bar{b}$ interactions, and $\kappa_{V,A}=1$ in the SM.

To probe the $Zb\bar{b}$ coupling using the WSSA measurement, a final state $b$-jet needed to be tagged in the DIS production.  As shown in Eq.~\eqref{eq:avq}, the average jet charge of a $b$-tagged jet,
$\langle Q_J\rangle_b(p_T^j)$, depends on the 
efficiency ($\epsilon_q^b$) of (mis-)tagging a $q$-jet as a $b$-jet. 
In this study, we adopt two sets of benchmark tagging efficiencies~\cite{Yan:2021htf} as 
\begin{align}
(i)~ & \epsilon_q^b=0.001,&&\epsilon_c^b=0.03,&&\epsilon_b^b=0.7;\nn\\
(ii)~ & \epsilon_q^b=0.01,&&\epsilon_c^b=0.2,&&\epsilon_b^b=0.5,
\label{eq:eff}
\end{align}
where $q=u,d,s$ denotes the light quarks. The scenario $(i)$  represents a good $b$-tagging efficiency, and $(ii)$ a worse one. 
After combining all possible quark flavor contributions, the $b$-tagged differential distribution can be approximated at the LO as
\beq
\frac{d \sigma_{b,\pm}^{\rm tot}}{dp_T^j}=\sum_{q=u,d,s}\frac{d\sigma_{q,\pm}}{dp_T^j}\epsilon_q^b+\frac{d\sigma_{c,\pm}}{dp_T^j}\epsilon_c^b+\frac{d\sigma_{b,\pm}}{dp_T^j}(\kappa_V,\kappa_A)\epsilon_b^b.
\label{eq:diffxsec}
\eeq

We will follow Ref.~\cite{Li:2019dre} and approximate the NLO inclusive jet cross section in the lepton-hadron collision as the product of the differential cross section of the production process 
${d\sigma_q(P_e,\mu_R,\mu_F)}/{dp_T^j}$  evaluated at the LO and the jet function ($J_q$) at the NLO . This approximation holds in the narrow-jet approximation when the needed matching coefficient is also calculated at the NLO, as explicitly shown in Ref.~\cite{Li:2019dre}. 
In our numerical calculation, we have set the energy of the electron ($E_e$) and proton ($E_p$) beams to be 18 GeV and 275 GeV, respectively. 
Both the renormalization and factorization scales are fixed at $\mu_R=\mu_F=\sqrt{Q^2+(p_T^j)^2}$, where $Q=\sqrt{-q^2}$ with $q^\mu$ being the momentum transfer of the electrons.
We have also required the final state jet (with jet size $R=1$) pass the following kinematical cuts: $p_T^j > 5$  GeV and $-2 < \eta_j < 4$, where $\eta_j$ is the rapidity of jet measured in the lab frame.
To calculate the heavy flavor contributions, such as the charm and bottom quarks, we adopt the sACOT-$\chi$ scheme~\cite{Aivazis:1993kh,Aivazis:1993pi,Collins:1998rz,Kramer:2000hn,Tung:2001mv}.
At the LO, this amounts to setting its mass ($m$) to be zero in calculating the scattering amplitude, while replacing the momentum fraction (Bjorken $x$) carried by the heavy parton by $\chi=x ( 1+ 4 m^2/Q^2)$~\cite{Tung:2001mv}. Here, we use $m_c=1.3$ GeV and $m_b=4.75$ GeV, to be consistent with the CT14LO parton distribution functions (PDFs)~\cite{Dulat:2015mca} adopted in this work. The strong coupling constant is taken to be $\alpha_s(M_Z)=0.118$ at the $Z$ boson mass scale.

Below, we give a detailed analysis of the dependence of $A_e^{bQ}$ on the $\kappa_V$ and $\kappa_A$ parameters. As shown in Ref.~\cite{Yan:2021htf}, $\kappa_V$ and $\kappa_A$ can contribute to the $b$-tagged SSA through  $g_A^e\cdot \kappa_V$ or $g_V^e\cdot \kappa_A$ in the $\gamma Z$ interference diagram, and  $\left((g_V^e)^2+(g_A^e)^2\right)\cdot \kappa_V\kappa_A$ or $g_V^eg_A^e\cdot(\kappa_V^2+\kappa_A^2)$ in the $Z$-only diagram, where $g_{V,A}^e$ are the vector and axial-vector components of the $Ze\bar{e}$ couplings. However, the contributions from $g_V^e\cdot \kappa_A$ and $\kappa_V\kappa_A$ 
are associated with the $F_3$ structure function, which 
involves the convolution of ($f_b - f_{\bar{b}}$), and must be vanishing at the LO and NLO, for the bottom quark PDF is generated perturbatively through DGLAP evolution~\cite{Moch:2004pa,Catani:2004nc}. 
Hence, up to the NLO, only the terms proportional to $g_A^e\cdot\kappa_V$ and $(\kappa_V^2+\kappa_A^2)$ will contribute to the $b$-tagged SSA.
On the other hand, the jet charged weighted cross section $\sigma_{b,\pm}^Q$ is weighted by the electric charge of $Q$, cf. Eq.~\eqref{eq:sigmabq}, as compared to the cross section $\sigma_{b,\pm}^{\rm tot}$ used in the calculation of SSA. Hence, on the contrary, the contributions from $g_A^e\cdot\kappa_V$ and $(\kappa_V^2+\kappa_A^2)$ in WSSA will be vanishing, and only the terms proportional to   $g_V^e\cdot\kappa_A$ and $\kappa_V\kappa_A$ will contribute to the $b$-tagged WSSA. Since the $Z$-only diagram will be suppressed by one more power of $Z$ propagator, we expect the WSSA would \textit{dominantly depend linearly on $\kappa_A$ through the $\gamma Z$ interference diagram}. Another important difference between WSSA and SSA is that the contribution from the $Z$-only diagram ($\kappa_V\kappa_A$ term) cannot be ignored in WSSA, because $g_V^e\ll g_A^e$. Similar to the SSA, the $\gamma$-only diagram will not change the conclusion and only contribute to the denominator of the WSSA.

\vspace{3mm}
\noindent{\bf Sensitivity at the EIC:} 
Below, we consider the WSSA measurement at the upcoming EIC with polarized electron beam.  For simplicity, we assume that both the right- and left-handed electron beams have the same  degree of polarization, with the same integrated luminosity.  The WSSA in Eq.~\eqref{eq:asy} can be related to the experimental observable as 
\beq
A_e^{bQ} = \frac{1}{P_e}
\frac{\sigma_b^{Q}(P_e)-\sigma_b^{Q}(-P_e)}{\sigma_b^{Q}(P_e)+ \sigma_b^{Q}(-P_e)},
\eeq
where $\sigma_b^{Q}(P_e)$ denotes the average jet charge weighted $b$-tagged cross section in the experiment  for the electron beam with polarization $P_e=70\%$ at the EIC~\cite{Accardi:2012qut}, and it  can be related to the cross sections with the incoming electron at its helicity eigenstates as,
\beq
\sigma_b^Q(P_e)=\frac{1}{2}\left(\sigma_{b,+}^Q+\sigma_{b,-}^Q\right)+\frac{P_e}{2}\left(\sigma_{b,+}^Q-\sigma_{b,-}^Q\right).
\eeq

\begin{figure}
	\includegraphics[scale=0.22]{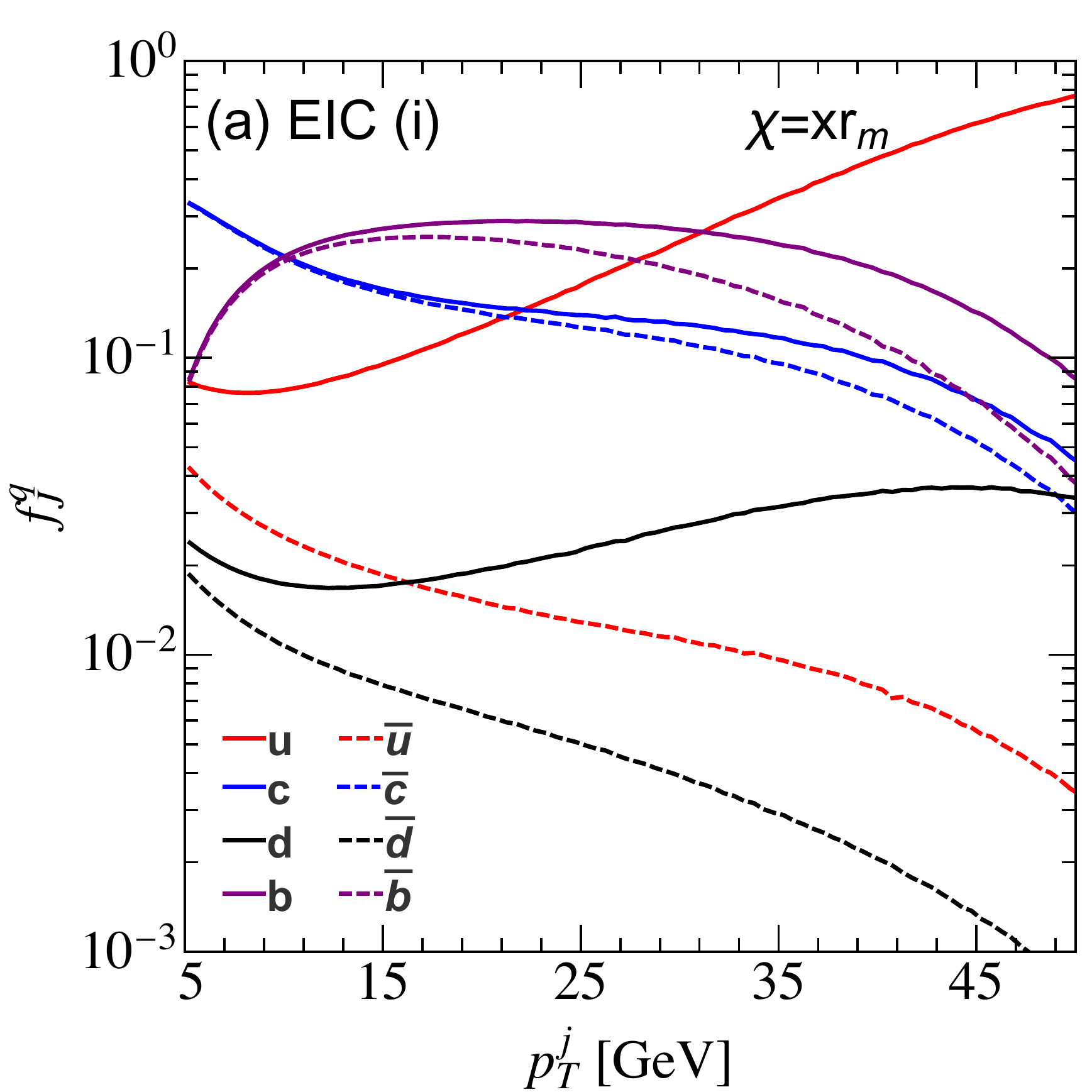}
	\includegraphics[scale=0.22]{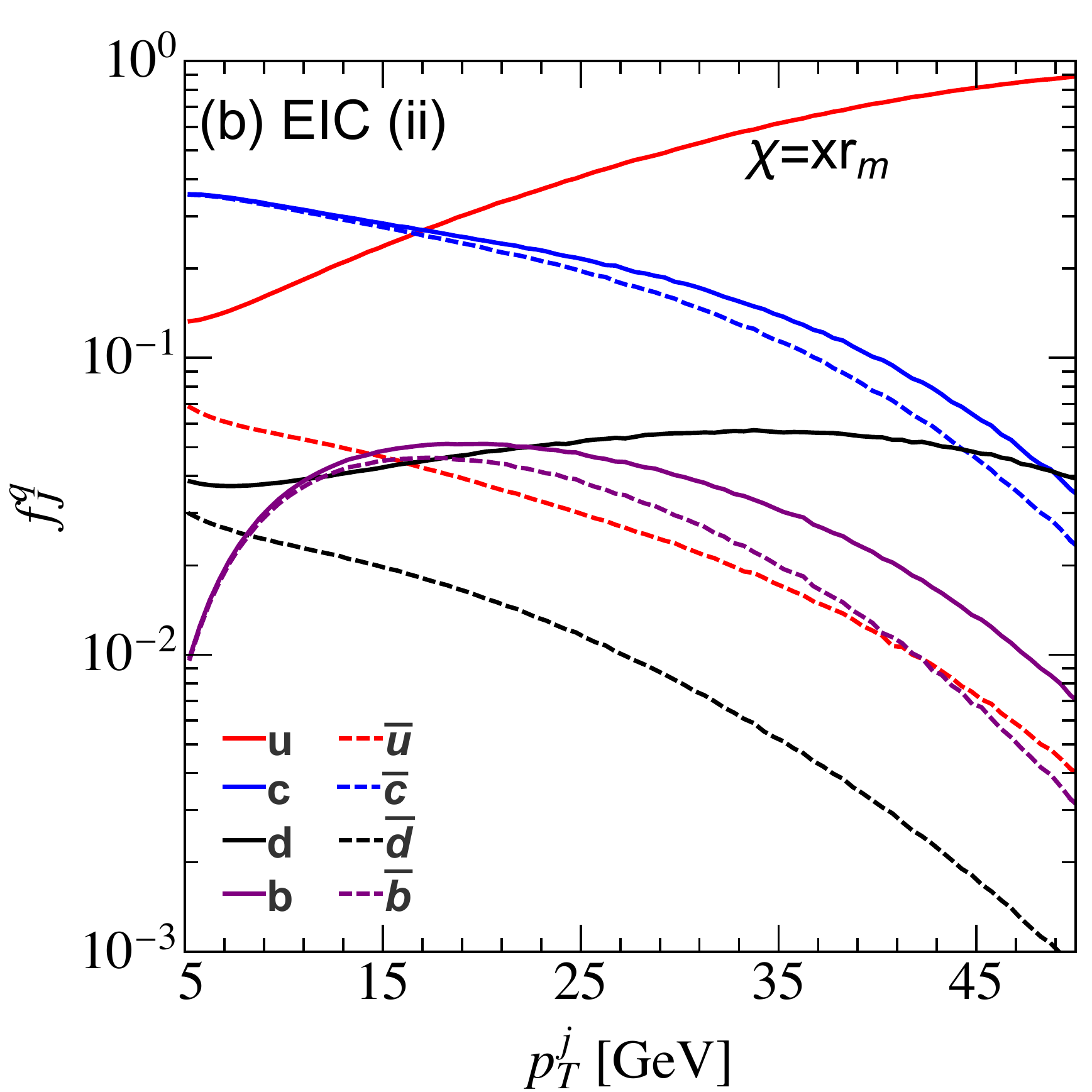}
	\includegraphics[scale=0.22]{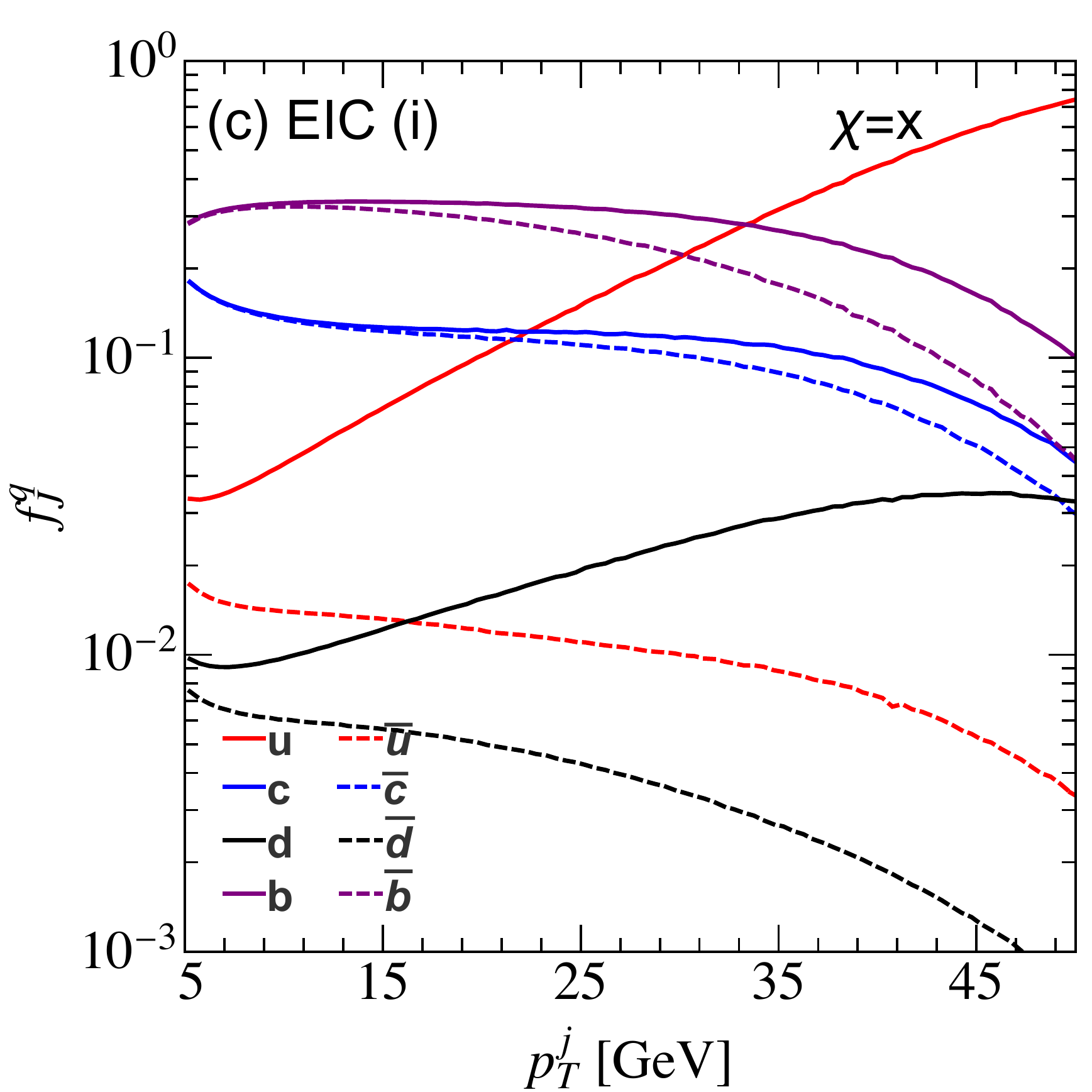}
	\includegraphics[scale=0.22]{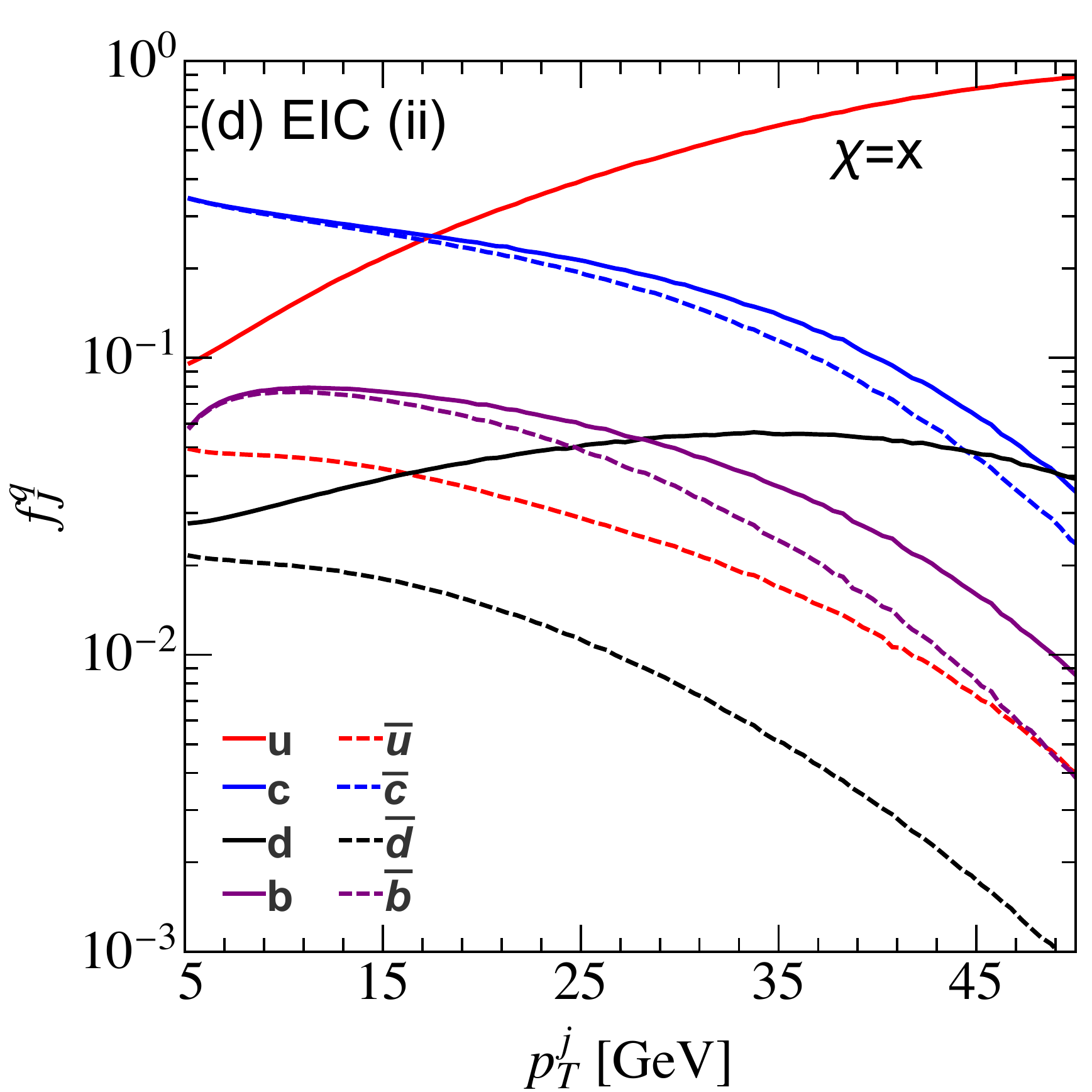}
	\caption{The quark fraction distributions $f_J^q$ as a function of jet $p_T^j$ at the EIC, with the tagging efficiency scenarios $(i)$ and $(ii)$, cf. Eq.~\eqref{eq:eff}, respectively.
Here, $x$ is the Bjorken-$x$ value, $r_m=1+4m_{c,b}^2/Q^2$, and the electron beam polarization $P_e=70\%$. Both the down and strange (anti-)quark contributions have been added in the $d$ and $\bar{d}$ fractions. }
\label{Fig:frac}
\end{figure}

\begin{figure}
	\includegraphics[scale=0.23]{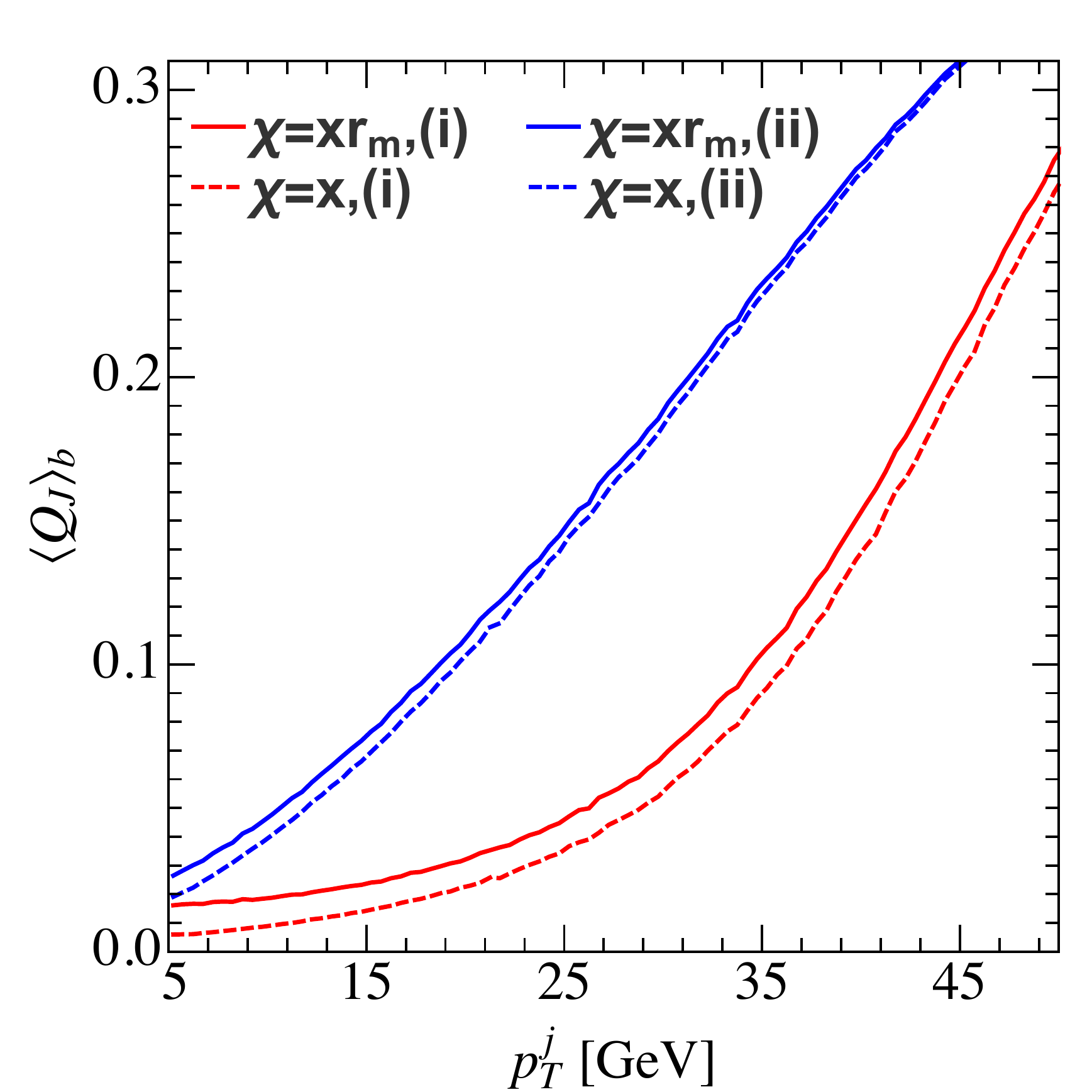}
	\caption{The total average jet charge distribution of a $b$-tagged jet $\langle Q_J\rangle_b$  
	as a function of jet $p_T^j$ at the EIC. 
	The red and blue curves correspond to the tagging efficiency scenarios $(i)$ and $(ii)$, cf. Eq.~\eqref{eq:eff}, respectively.  Here, $x$ is the Bjorken-$x$ value, $r_m=1+4m_{c,b}^2/Q^2$, and the electron beam polarization $P_e=70\%$.} 
\label{Fig:jQtot}
\end{figure}

\begin{figure*}
	\includegraphics[scale=0.23]{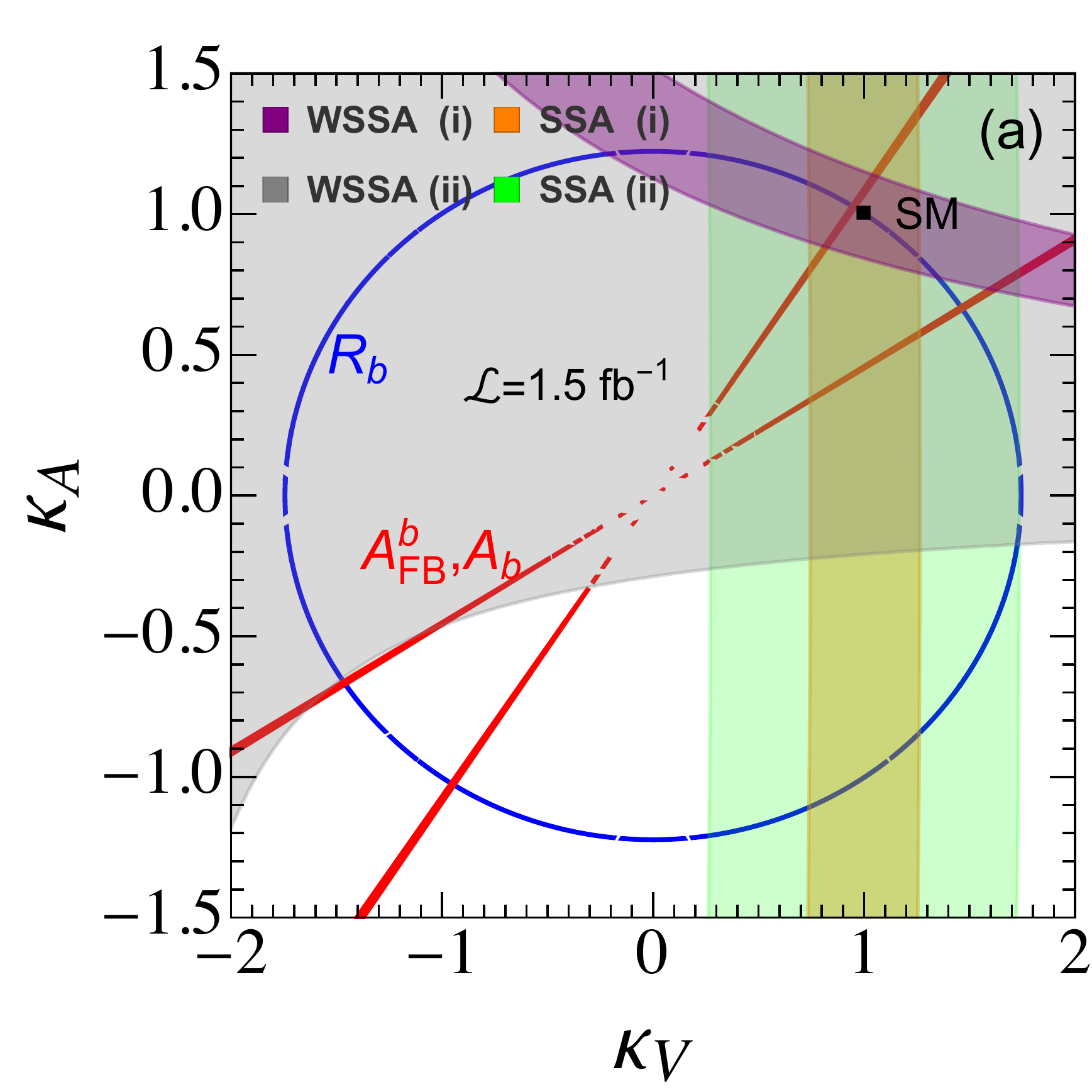}
	\includegraphics[scale=0.23]{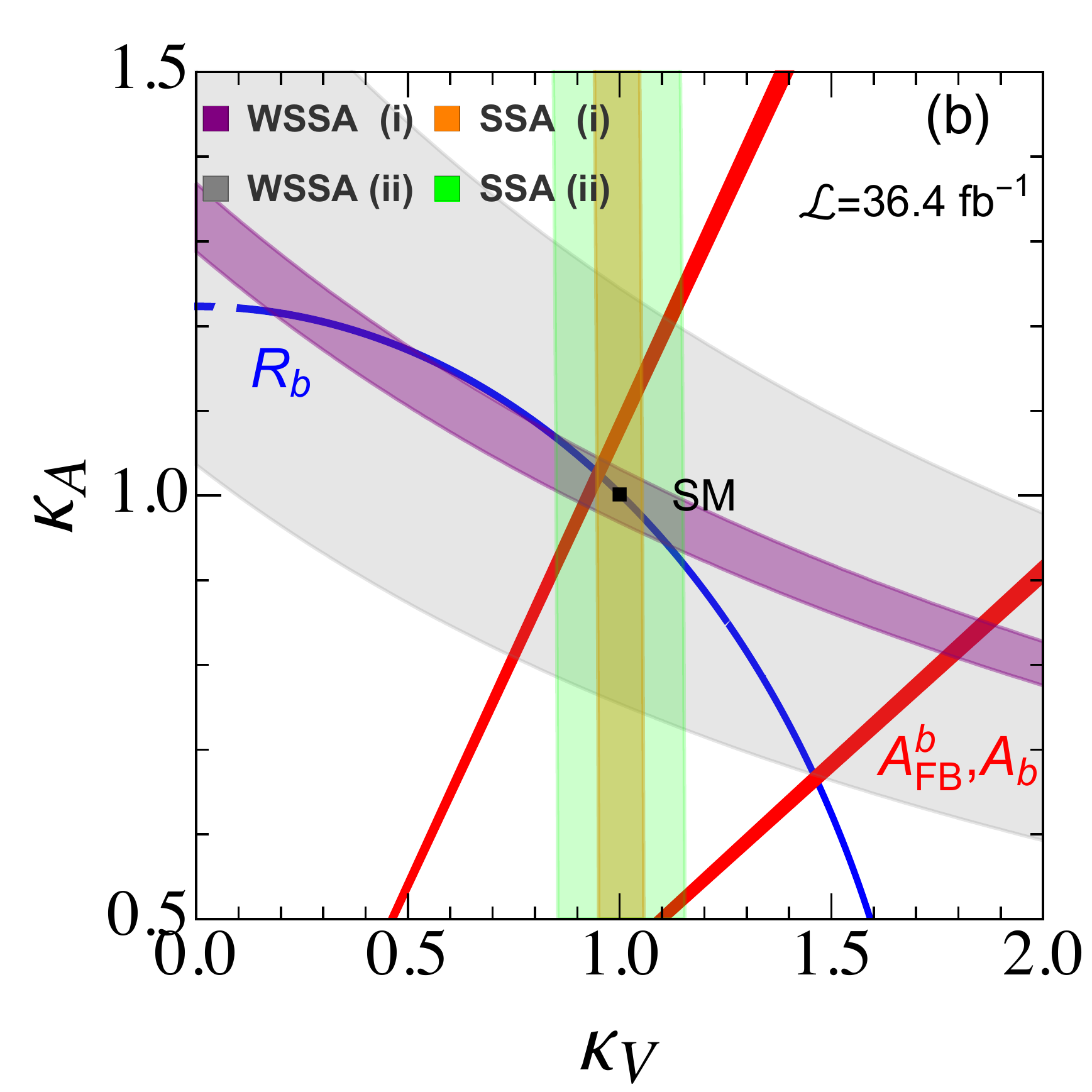}
	\includegraphics[scale=0.23]{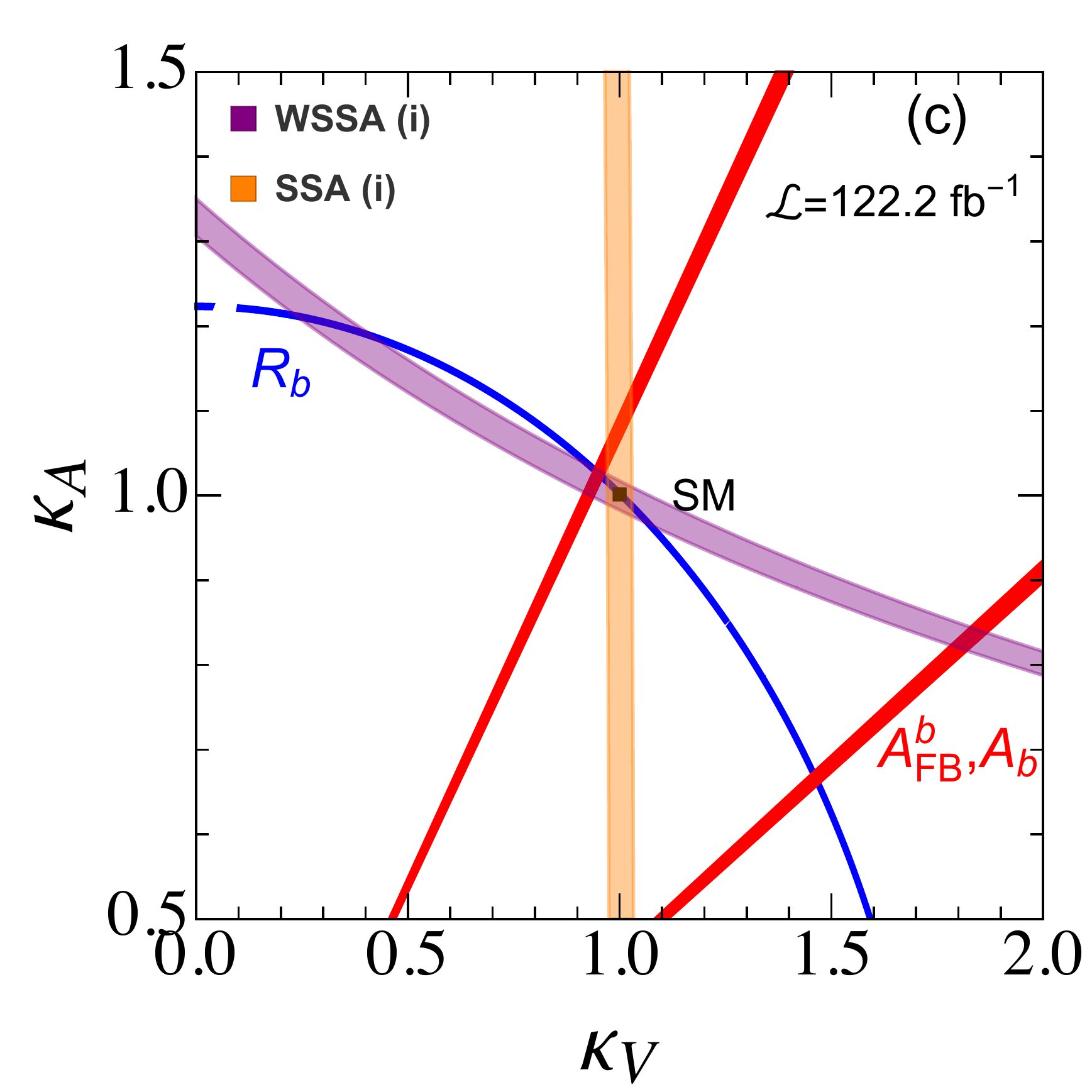}
	\caption{
The expected limits on the $Zb\bar{b}$ anomalous couplings $\kappa_V$ and $\kappa_A$ at the EIC with 68\% C.L. The blue and red regions come from the $R_b$ and $(A_{\rm FB}^b, A_b)$ measurements at the LEP and SLC, respectively. The purple and gray bands come from the measurements of the WSSA $A_{e}^{bQ}$ at the EIC with the tagging efficiency scenarios $(i)$ and $(ii)$, cf. Eq.~\eqref{eq:eff}, respectively, while the orange and green shaded regions correspond to the expected results from the SSA measurement.
}
	\label{Fig:EIC}
\end{figure*}

With the considered tagging efficiencies in Eq.~\eqref{eq:eff} and the degree of polarization of the electron beam at the EIC, we calculate various (anti-)quark fraction distributions $f_J^q$, 
which is defined as:
\beq
f_J^q(p_T^j,\epsilon_q^b)\equiv 
\Big({d\sigma_q \over dp_T^j} \epsilon_q^b \Big)\Big / \Big({d\sigma_b^{\rm tot}\over dp_T^j}\Big),
\eeq
where the differential cross section ${d\sigma_q(P_e,\mu_R,\mu_F)/ dp_T^j}$, cf.  
Eq.~\ref{eq:diffxsec},
can be written as a convolution of the PDFs and partonic cross section according to the collinear factorization theorem.  
Note that both the down and strange (anti-)quark contributions have been added in $f_J^{d}$ and $f_J^{\bar{d}}$, for simplicity. With the result depicted in Fig.~\ref{Fig:frac}, several comments are in order:
\begin{itemize}
\item The fractions $f_J^{u(d)}\gg f_J^{\bar{u}(\bar{d})}$, while $f_J^{c(b)} \sim f_J^{\bar{c}(\bar{b})}$, because the up and down quarks contain both the valence and sea quarks (with $u_{sea}=\bar u$, etc.), while the charm and bottom quarks are sea quarks.
\item We notice a small difference between $f_J^{c(b)}$ and $f_J^{\bar{c}(\bar{b})}$ in the high $p_T^j$ region. It arises from the fact that the axial-vector  $Zq\bar{q}$ coupling yields an opposite contribution from quark and anti-quark scattering processes~\cite{Yan:2021htf}. The difference becomes obvious only when the interference effect induced by the $\gamma Z$ diagram leads to a sizable contribution to the total cross section, {\it i.e.,} in the high $p_T^j$ region.
\item The quark fraction distributions $f_J^q$ are strongly dependent on the jet tagging efficiencies, because the cross sections of the light quarks  are much larger than the bottom quark.
\item The fractions $f_J^{b,\bar{b}}$ are very sensitive to the bottom quark mass in the small $p_T^j$ region due to the rescaling of the Bjorken $x$ in the sACOT-$\chi$ scheme adopted in this calculation.
\end{itemize}

In Fig.~\ref{Fig:jQtot}, we show the total average jet charge distributions, cf.  Eq.~\eqref{eq:avq}, with $\kappa=0.3$ at the EIC. The red and blue curves denote the predictions with the tagging efficiency scenarios $(i)$ and $(ii)$ in Eq.~\eqref{eq:eff}, respectively. It shows that the expected $\langle Q_J \rangle_b$ is sensitive to the jet tagging efficiencies and the heavy quark mass effects. Such behavior could be understood from the quark fraction distributions $f_J^q$ in Fig.~\ref{Fig:frac}. The comparison of Figs.~\ref{Fig:frac}(c) and (d) shows that the fraction $f_J^u$ is significantly enhanced with the tagging scenario $(ii)$, as compared to $(i)$, while this enhancement is suppressed after including the heavy quark mass effects via adopting the sACOT-$\chi$ scheme, cf. Figs.~\ref{Fig:frac}(a) and (b). Also, the behavior of $f_J^{b,\bar{b}}$ is different from that of $f_J^u$.

Below, we combine the differential distribution $d\sigma_{b,\pm}^{\rm tot}/dp_T^j$ and the total average jet charge to estimate the sensitivity of the WSSA to the $Zb\bar{b}$ anomalous couplings at the EIC. The systematic uncertainties are assumed to be cancelled in the WSSA definition and will be ignored in this work~\cite{Chekanov:2009gm}. 
With the tagging efficiencies $(i)$ and $(ii)$ in Eq.~\eqref{eq:eff}, we obtain the WSSA at the EIC,
\begin{align}
(i)&\quad A_e^{bQ}=\frac{10.3-13.0\kappa_A-4.6\kappa_V\kappa_A}{-1073.9+117.2\kappa_A+\kappa_A\kappa_V},\nn\\
(ii)&\quad A_e^{bQ}=\frac{120.5-13.0\kappa_A-4.6\kappa_V\kappa_A}{-14748.5+117.2\kappa_A+\kappa_A\kappa_V}.
\end{align}
It shows that $A_e^{bQ}$ only depends on the linear combination of $\kappa_A$ and $\kappa_V\kappa_A$, as expected, with a larger coefficient associated with the former term.
The statistical uncertainty of $A_e^{bQ}$ could be well controlled due to the large DIS cross section, i.e. $(i)$ $\sigma_b^{\rm tot}=77~{\rm pb}$ and $(ii)$ $\sigma_{b}^{\rm tot}=417~{\rm pb}$. 
With the integrated luminosity $\mathcal{L}=100~{\rm fb}^{-1}$,  the WSSA in the SM are $(i)$ $A_e^{bQ}=0.008$, with $\delta A_e^{bQ}/A_e^{bQ}=5\%$; $(ii)$ $A_e^{bQ}=-0.007$, with $\delta A_e^{bQ}/A_e^{bQ}=3\%$, where $\delta A_e^{bQ}$ denotes the statistical uncertainty of $A_e^{bQ}$. We notice that $A_e^{bQ}$ changes sign from the case with tagging efficiency $(i)$ to $(ii)$. It arises from the fact that the jet charge weighted cross sections of the non-$b$ jets are highly enhanced when using a worse tagging efficiency, as specified in $(ii)$.
 
In Fig.~\ref{Fig:EIC}, we show the expected limits, at the 68\% confidence level (C.L.), on the anomalous $Zb\bar{b}$ couplings for various integrated luminosities, obtained from the WSSA (purple and gray bands) and SSA (orange and green bands) measurements at the EIC. The blue and red bands denote the limits imposed by the $R_b$ and $(A_{\rm FB}^b,A_b)$ measurements at the $Z$-pole, respectively. It is evident that the measurement of $A_e^{bQ}$ ( purple and gray bands) is more sensitive to the axial vector component of the $Zb\bar{b}$ coupling, {\it i.e.}, $\kappa_A$, while the SSA (orange and green bands) is more sensitive to the vector component $\kappa_V$, cf.  Ref.~\cite{Yan:2021htf}.  The shapes of the bands from WSSA are determined by the relative size of the $\gamma Z$ and $Z$-only diagrams and dominantly depend on the  interference effect of the $\gamma Z$ diagram. We note that  the WSSA and SSA have a different sensitivity to the $Zb\bar{b}$ couplings as compared to the measurement of $gg\to Zh$ scattering cross section at the LHC~\cite{Yan:2021veo}.  Therefore, the measurements of WSSA and SSA at the EIC provide complementary information on the $Zb\bar{b}$ couplings to the above-mentioned measurements conducted at the LHC and lepton colliders.

We could also derive the minimal amount of integrated luminosities needed to exclude the degeneracy parameter space, implied by the precision electroweak data, with $\kappa_{V,A}<0$ at the 68\% C.L. The results for the two choices of the tagging efficiencies ($i$ and $ii$ in Eq.~\eqref{eq:eff}) are found to be (see Fig.~\ref{Fig:EIC}(a)) 
\beq
(i): \mathcal{L}>0.03~{\rm fb}^{-1};\quad\quad (ii): \mathcal{L}>1.5~{\rm fb}^{-1}.
\eeq

To resolve the apparent degeneracy in the parameter space with $\kappa_{V,A}>0$, {\it i.e.,}  ($\kappa_V,\kappa_A$)=(1.46,0.67), the needed minimal luminosities are  (see Fig.~\ref{Fig:EIC}(b))
\beq
(i): \mathcal{L}>0.6~{\rm fb}^{-1};\quad (ii): \mathcal{L}>36.4~{\rm fb}^{-1}.
\eeq
Finally, the minimal luminosities to exclude the LEP $A_{\rm FB}^b$ measurement (the solution which is close to the SM) through WSSA are (see Fig.~\ref{Fig:EIC}(c))
\beq
(i): \mathcal{L}>122~{\rm fb}^{-1};\quad (ii): \mathcal{L}>8583~{\rm fb}^{-1}.
\eeq
Note that the constraint with the worse tagging efficiency $(ii)$ is not plotted in Fig.~\ref{Fig:EIC}(c), for the required luminosity to exclude the $A_{FB}^b$ anomaly is too large.
Although it would be challenging to verify or exclude the LEP $A_{\rm FB}^b$ measurement through WSSA with a worse tagging efficiency, such goal could be realized via SSA over a few years of running as shown in Ref.~\cite{Yan:2021htf}.

\vspace{3mm}
\noindent {\bf Conclusions:~}%
In this work, we propose a novel method to probe the $Zb\bar{b} $ anomalous couplings by measuring the average jet charge weighted single-spin asymmetry $A_e^{bQ}$, in the total inclusive $b$-tagged DIS cross section, of a polarized electron beam scattering off an unpolarized proton beam at the EIC. We show that both the quark fraction $f_J^q(p_T^j)$ and the average jet charge
$\langle Q_J\rangle_b(p_T^j)$ 
distributions of a $b$-tagged jet  
are sensitive to the $b$-tagging efficiency and the mass of heavy quarks. As a result, the determination of the $Zb\bar{b}$ couplings from WSSA is sensitive to the $b$-tagging efficiency.
We demonstrated that the WSSA  $A_e^{bQ}$ is sensitive to the axial-vector component of the $Zb\bar{b}$ coupling, which is in contrast to the sensitivity to the $Zb\bar{b}$ vector component from the SSA measurement and plays a complementary role on constraining the $Zb\bar{b}$ anomalous couplings. Similar to the SSA, the WSSA at the EIC could also clarify the long-standing discrepancy between the LEP $A_{\rm FB}^b$ measurement and the SM prediction owing to the high luminosity of the EIC.

\vspace{3mm}
\noindent{\bf Acknowledgments.}
This work is partially supported by the U.S. Department of Energy, Office of Science, Office of Nuclear Physics, under Contract DE-AC52-06NA25396 through the LANL/LDRD Program, as well as the U.S.~National Science Foundation
under Grant No.~PHY-2013791. C.-P.~Yuan is also grateful for the support from the Wu-Ki Tung endowed chair in particle physics.

\bibliographystyle{apsrev}
\bibliography{reference}

\end{document}